# Deep Learning Fundus Image Analysis for Diabetic Retinopathy and Macular Edema Grading

Jaakko Sahlsten[1], Joel Jaskari[1], Jyri Kivinen[1], Lauri Turunen[2], Esa Jaanio[2], Kustaa Hietala[3] & Kimmo Kaski[1,*]

February 18, 2019

[1]Dept. of Computer Science, Aalto University School of Science, 00076, Finland. [2]Digifundus Ltd., Tietotie 2, 90460 Oulunsalo, Finland. [3]Central Finland Central Hospital, Keskussairaalantie 19, 40620 Jyväskylä, Finland. *Corresponding author: e-mail: kimmo.kaski@aalto.fi, regular mail: P.O. Box 15500, FI-00076 AALTO, Finland.

## Abstract

Diabetes is a globally prevalent disease that can cause visible microvascular complications such as diabetic retinopathy and macular edema in the human eye retina, the images of which are today used for manual disease screening. This labor-intensive task could greatly benefit from automatic detection using deep learning technique. Here we present a deep learning system that identifies referable diabetic retinopathy comparably or better than presented in the previous studies, although we use only a small fraction of images (<1/4) in training but are aided with higher image resolutions. We also provide novel results for five different screening and clinical grading systems for diabetic retinopathy and macular edema classification, including results for accurately classifying images according to clinical five-grade diabetic retinopathy and four-grade diabetic macular edema scales. These results suggest, that a deep learning system could increase the cost-effectiveness of screening while attaining higher than recommended performance, and that the system could be applied in clinical examinations requiring finer grading.

## Introduction

Diabetic retinopathy is the most common microvascular complication in diabetes[1], for the screening of which the retinal imaging is the most widely used method due to its high sensitivity in detecting retinopathy[2]. The evaluation of the severity and degree of retinopathy associated with a person having diabetes, is currently performed by medical experts based on the fundus or retinal images of the patient's eyes[3]. As the number of patients with diabetes is rapidly increasing, the number of retinal images produced by the screening programmes will also increase, which in turn introduces a large labor-intensive burden on the medical experts as well as cost to the healthcare services. This could be alleviated with an automated system either as support for medical experts' work or as full diagnosis tool. There are two recent studies that have investigated the use of deep learning systems in automated detection of diabetic retinopathy[4,5]. Both show that an automated system, based on the deep learning artificial neural network approach, can achieve high sensitivity with high specificity in detecting the referable diabetic retinopathy, defined as moderate or worse diabetic retinopathy. There are also other referable eye complications that have recently been investigated with this approach, such as diabetic macular edema[4], and possible glaucoma and age-related macular degeneration[5].

For an automated system to be clinically viable, it should be able to classify retinal images based on clinically used severity scales, such as the proposed international clinical diabetic retinopathy and diabetic macular edema disease scales[6], which are also used in Finland[7]. In the literature one can find recent experiments[8,9] for the former case of diabetic retinopathy scale, but there are no experiments yet to classify macular changes with the latter scale. Another substantial barrier to broader and more effective use of deep learning system is thought to be the large quantity of annotated images needed for the model to learn[10].

In this study, our aim is to identify retinopathy using five different diabetic retinopathy classification systems. Moreover, we present what preprocessing and regularization steps to the images needs to be done for the good functionality of the deep learning system and investigate systematically how the size with much smaller number of images used in training affects its performance.

## Methods

### Original fundus image dataset

The research of present study was done in collaboration with Digifundus Ltd, an ISO 9001:2015 certified provider of diabetic retinopathy screening and monitoring services in Finland. Digifundus Ltd provided a non-open, anonymized retinal image dataset of patients with diabetes, including 41122 graded retinal color images from 14624 patients. The images were taken with Canon CR2 retinal camera after inducing mydriasis with tropicamide 5 mg/ml eyedrops. Two 45 degree color fundus photographs, centered on fovea and optic disc were taken from the patient's both eyes. The output images were of variable resolutions, ranging from $3888 \times 2592$ to $5184 \times 3456$ pixels.

The present study is a methodological study with anonymized medical data and without any intervention in the integrity of a person such as contact with a person. In Finnish law this is not considered as a medical study requiring approval by an ethics committee or a written consent of a person[11].

### Retinal image grading systems and gradability

Each of the retinal images had been graded with respect to three different criteria, i) diabetic retinopathy, ii) macular edema, and iii) gradability. Images are graded with the proposed international clinical diabetic retinopathy and macular edema disease severity scales[6], denoted later as PIRC and PIMEC, respectively. The image gradability is a two-stage system, which considers an image to be either gradable or not. All personnel participating in retinopathy assessment had over 10 years' experience in diabetic retinopathy grading. Retinal images with no lesions or mild diabetic lesions were graded by an optometrist and an M.D. trained for retinopathy grading. All images with moderate or worse changes were graded by two ophthalmologist both with more than 10 years of experience in grading fundus images. If there was a disagreement in grading, such an image was not included in this study.

PIRC and PIMEC grades were further used to obtain additional three types of grading systems: i) a binary system of *nonreferable*/*referable diabetic retinopathy* (NRDR/RDR), ii) a binary system of *nonreferable*/*referable diabetic macular edema* (NRDME/RDME), and iii) three-class system of ungradable/NRDR/RDR. The NRDR/RDR system considers the cases with no diabetic retinopathy and mild diabetic retinopathy as nonreferable diabetic retinopathy, and the cases with moderate or worse diabetic retinopathy as referable diabetic retinopathy. This system has been used in recent works investigating automated detection of diabetic retinopathy[4,5]. The NRDME/RDME system here is defined such that the absence of macular edema is defined as nonreferable diabetic macular edema and any level of macular edema as referable diabetic macular edema. Note that only the gradable images were graded for diabetic retinopathy and macular edema. Ungradable images were included in a single task, in combination with referable diabetic retinopathy classification, which constitutes the grading system QRDR, in which each image is considered to be either ungradable, depicting non-referable diabetic retinopathy, or depicting referable diabetic retinopathy (ungradable/NRDR/RDR).

## Image preprocessing and dataset division

In the model training and subsequent primary validation, we used preprocessed versions of the original images. The preprocessing consisted of image cropping followed by resizing. Each image was cropped to a square shape which included the most tightly contained circular area of fundus. The procedure removed most of the black borders and all of the patient related annotations from the image data. Each of the cropped images were then resized to five different standard input image sizes of $256 \times 256$, $299 \times 299$, $512 \times 512$, $1024 \times 1024$, and $2095 \times 2095$ pixels. Here the creation of multiple resolutions was done for the purposes of analyzing the effect of the input image resolution on the classification performance.

The obtained processed datasets were divided into three sets: *training*, *tuning*, and *primary validation* set in the 70 %, 10 % and 20 % proportions of the whole image dataset, respectively, separately for each of the grading systems used in the experiments. In the division per a particular grading system, the different sets were to have similar grade distributions, and that the dataset data per patient to not reside in multiple but only in one of the three sets (of training, tuning, and primary validation), in order to prevent the possibility of obtaining over-optimistic results due to data memorization. Table 1 shows the statistics of the resulting divisions that were used in the experiments. Note that the grade distributions across the different sets were similar, with respect to each grading system, for example, when we consider the NRDR/RDR-system, the proportion of images associated with referable diabetic retinopathy in the training, tuning and primary validation set 44 %, 43.9 % and 43.4 %, respectively.

## Deep learning model

In order to distinguish features related to diabetic retinopathy and macular edema in the color images of patients' fundi we chose to use a deep convolutional neural network. The neural networks are mathematical models, which consist of parameters used in specific calculations, such as convolutions and summations. The neural network can be constructed in such a way, that it receives an input which is used in calculating an output[12], such as class or grade of diabetic retinopathy. The parameters used in the calculation of the output can be modified in a data-driven manner, by minimizing the error between neural network produced detections of classes and the manual annotations.

The network architecture, we selected, was based on the Inception-v3 architecture[13], pretrained on ImageNet dataset[14], similar to the base model used for the classification of referable diabetic retinopathy in Gulshan et al.[4]. A detailed description of the neural network training and design is presented in the in the Supplementary information.

## Model evaluation and comparison against previous works

The present study was conducted by training a separate model for each of the five classification tasks and five input image sizes. To evaluate the performance of our model in binary classification tasks we use the receiver operating characteristic (ROC) curve from which we determine the area under the ROC curve (AUC) as well as accuracy, sensitivity, and specificity, while in the multi-class cases we use the area under macro average of ROC (macro-AUC) for each class calculated in one-vs-all manner, accuracy and quadratic-weighted kappa score. Also, we calculate the confusion matrices for the multi-class classification tasks. For each metric in the binary classification tasks, the exact 95% confidence interval (CI) was calculated using the Clopper-Pearson method, similar to that in Gulshan et al.[4] for comparison. In a recent publication[5] a different confidence interval estimation method was used, namely the cluster-bootstrap, providing a bias-corrected, asymptotic 2-sided 95% CI. This approach used a patient-level clustering for estimating the AUC at the population level.

We conduct the performance comparison of our model against the performance of the recently presented systems described by Gulshan et al.[4], Ting et al.[5] and in Krause et al.[8] The study by Gulshan et al.[4] using a large set of labeled images, was the first deep learning system achieving high values of sensitivity and specificity in detecting referable diabetic retinopathy. The system developed by Ting et al.[5] for the same task, was trained using large datasets of multi-ethnic population, achieved nearly comparable results to those by Gulshan et al.[4] A more recent work described in Krause et al.[8] extended these works, by focusing on classifying the severity of diabetic retinopathy using the five-level PIRC scale. The present study extends these previous studies: (i) by using significantly less images to train the deep learning system; (ii) by conducting experiments on five different classification systems, including two clinically used scales, PIRC and PIMEC; (iii) by performing training for different tasks without aggregated model prediction to provide independent results for each task; and (iv) by investigating the effect of image resolution to the model, in order to find out their effect on model performance and a trade-off between the number (or cost) of manually annotated images, image resolution and performance.

## Results

In the binary classification tasks, i.e. NRDR/RDR and NRDME/RDME, our algorithm achieved the best results using the largest 2095 × 2095 pixels input image size. In the NRDR/RDR classification on our primary validation set having 7118 images, our algorithm achieved the sensitivity of 0.896 (with 95% CI: 0.885-0.907) and specificity 0.974 (with 95% CI: 0.969-979) and AUC of 0.987 (with 95% CI: 0.984-0.989). Our model performance was evaluated at the operating point where the tuning set achieved 0.900 sensitivity, in a similar manner to Ting et al[5], while Gulshan et al.[4] had two operating points namely at a high specificity (0.980) point and at a high sensitivity (0.975) point. In Table 2 we present the AUC values of our model, along with the AUC values reported by Gulshan et al.[4] and Ting et al.[5]. The Table 2 also illustrates our results and the results reported by Ting et al.[4] at 0.900 sensitivity operating point and results reported by Gulshan et al.[4], closest to the 0.900 sensitivity operating point. Two other recent studies, Krause et al.[8] and Guan et al.[9], also explored the NRDR/RDR classification, but as they do not report results close to the 0.900 sensitivity point, we make a separate comparison with these studies.

As we can see from the Table 2, our result for the AUC, as determined from the ROC curve (see Figure 1), and the specificity on a similar sensitivity point are on par with the results by Gulshan et al.[4]. Our model outperforms the system proposed by Ting et al.[5] in AUC and specificity at the same sensitivity point, 0.900. We have also experimented with a similar operating point selection as Gulshan et al.[4] When operating point was selected for clinical setting, having high specificity of approximately 0.980, our model scored 0.883 (0.880-0.886) sensitivity and 0.980 (0.979-0.981) specificity, in comparison to 0.903 (0.875-0.927) sensitivity and 0.981 (0.978-0.985) specificity reported in Gulshan et al.[4] With the operating point chosen for screening, having high sensitivity, our model scored 0.968 (0.961-0.974) sensitivity and 0.893 (0.883-0.902) specificity, compared to 0.975 (0.958-0.987) sensitivity and 0.934 (0.928-0.940) specificity reported by Gulshan et al.[4]

The study presented in Guan et al.[9] also considers the problem of classifying the stage of diabetic retinopathy as non-referable vs. referable diabetic retinopathy (NRDR/RDR). However, the method they proposed differs considerably from ours, as it models multiple labels, one for each individual labeler (medical expert), whereas our labels are constructed based on agreement between multiple labelers such that each model predicts only one label. In their case the best performing model is trained to predict the PIRC label for an image, and NRDR/RDR label is obtained by aggregating the PIRC grade probabilities. The best results presented in their work include 0.9745 AUC and an 0.9093 accuracy for the NRDR/RDR binary classification task, which are both outperformed by our best results of 0.987 AUC and an accuracy of 0.940. They also present 0.8361 specificity at 0.97 sensitivity point, which was outperformed by our model at a similar point with 0.960 sensitivity and 0.909 specificity. The dataset used in Guan et al.[9] was reported to be the same as used by Gulshan et al.[4], however the reported input image size was

587 × 587 pixels.

The more recent study presented in Krause et al.[8] also considers NRDR/RDR classification using aggregated PIRC predictions. They report the AUC of their best model being 0.986, which is comparable to our 0.987 AUC. They also presented the specificity of 0.923 at 0.971 sensitivity, which outperforms our models specificity of 0.909 at 0.960 sensitivity value.

The highest AUC for detecting referable diabetic macular edema (RDME) achieved by our model was 0.989 (0.986-0.991) with 2095 × 2095 resolution images. In Gulshan et al.[4] and Krause et al.[8], the NRDME/RDME classification performance is reported at specified operating points. In Gulshan et al.[4] an operating point with 0.908 (0.861-0.943) sensitivity and 0.987 (0.984-0.990) specificity was reported, whereas at a similar point sensitivity of 0.905 (0.887-0.922) our model had specificity of 0.978 (0.974-0.982), slightly underperforming in comparison. In Krause et al.[8] the sensitivity of 0.949 and specificity of 0.944 were reported for the detection of RDME. Our model achieved 0.954 specificity at similar sensitivity point of 0.947, thus slightly outperforming their proposed system.

In the multiclass classification tasks, i.e. PIRC, PIMEC, and QRDR, our algorithm achieved the best results, in terms of the macro-AUC, using 2095 × 2095 pixels input image size for PIRC and PIMEC tasks and for QRDR best results were achieved with 1024 × 1024 and 2095 × 2095 image sizes. In the PIRC classification task on our primary validation set having 7129 images, our algorithm achieved best performance measured in macro-AUC value of 0.962, with accuracy of 0.869 and quadratic weighted kappa of 0.910. When the performance was measured using quadratic-weighted kappa, the best performing model was achieved using 1024 × 1024 pixels input image size, with macro-AUC value of 0.961, accuracy of 0.870 and quadratic-weighted kappa of 0.915. Krause et al.[8] reported their system for PIRC classification having the quadratic-weighted kappa value of 0.84, which is outperformed by all our models trained by using larger than 299 × 299 pixels input image size models. Results for multiclass classifications tasks for each image size are shown in Table 3.

It should be noted that our model was trained using only 24 % (28512) of the number of images used by Gulshan et al.[4] (118419), 37 % of the number of images used by Ting et al.[5] (76370) and only 1.7% of images used by Krause et al.[8] (1662646). However, we trained and evaluated different models using multiple image sizes ranging from 256 × 256 pixels to 2095 × 2095 pixels, whereas the main results reported by Gulshan et al.[4], Ting et al.[5] and Krause et al.[8] were for image sizes 299 × 299 pixels, 512 × 512 pixels and 779 × 779 pixels, respectively. To summarize, the results demonstrate that our proposed system obtained comparable performance to those of the state-of-the-art systems, but by using considerably less or even a small fraction of training images.

In the QRDR task on the primary validation set of 8226 images, our algorithm achieved the best results in macro-AUC, accuracy and quadratic-weighted kappa, using the 1024 × 1024 pixels input image size. The model achieved macro-AUC of 0.991, accuracy of 0.938 and quadratic-weighted kappa of 0.932. In the PIMEC classification task the primary validation set of 7304 images, our algorithm achieved the best results for the macro-AUC and quadratic-weighted kappa using 2095 × 2095 resolution images and the best results for accuracy using 1024 × 1024 resolution images. The model with the greatest macro-AUC had the value of 0.981 with accuracy of 0.934 and quadratic-weighted kappa of 0.856. The ROC curves of the best performing models based on macro-AUC are shown in Figure 2.

The effect of the input image size on the ROC curves in binary classification tasks are illustrated in Figure 1, in which NRDR/RDR and NRDME/RDME classification systems are considered. From the subfigures and the tables we can see that increasing the input image resolution from 256 × 256 to 512 × 512 clearly improved the results, and even better results were obtained as the resolution was further increased; note that the modeling setup was a bit different for the 2095 × 2095 sized input images than for the others, as mentioned earlier. A similar observation was also made by Krause et al.[8], however their analysis was constrained in the context of PIRC classification, whereas we have systematically analyzed the effect of image resolution for five different classification systems.

In Figure 1 we provide illustrations of the ROC curves on the Messidor dataset[14]. From these curves we can see that the dependence of the model performance on the input image size is not as clear for the Messidor dataset. This is possibly in part due to the fact that Messidor images have multiple different resolutions, the smallest being of 900 × 900 pixels, which were resized into the sizes shown in Figure 1. In addition, the imaging equipment used in capturing the Messidor dataset and out primary dataset differ, and it is reported by Decencière et al.[15] that one third of the Messidor images were captured without Tropicamide induced pupil dilation. The detailed numerical results for each input image size on the multi-class and binary classification tasks are presented in Tables 3, 4 and 5. Confusion matrices for multiclass tasks are shown in Table 6.

It should be noted that a different larger Messidor set, called Messidor-2, was used by Gulshan et al.[4] than the standard Messidor set being only a subset of Messidor-2, for which reason our results are not directly comparable. In addition, it was reported that Messidor-2 dataset was annotated for NRDR/RDR system in Gulshan et al.[4], whereas our labels were constructed from the provided Messidor labels to be as close as possible counterparts of the NRDR/RDR system. The labels used in this study are thus not guaranteed to have translated correctly.

## Discussion

In this study, we have presented a systematic computational methodology for diabetic retinopathy classification, and assessed its performance on a non-open dataset using five different diabetic retinopathy classification systems. We have found that our deep learning model achieved comparable or better results with only a small fraction (<1/4) of training set images than used recently by two other groups to obtain the state-of-the-art results in the non-referable/referable diabetic retinopathy (NRDR/RDR) classification, with similar model architecture. The goodness of these results can most likely be attributed on one hand to regularizing image preprocessing and on the other hand to the features in the dataset and in the experimental setting. For example, our database was prepared with class/grade-balance in mind, so that its grade-distribution when considering the NRDR/RDR classification, was aimed to be uniform thus having grade distribution which does not necessarily follow a population or a clinical distribution. Other attributing aspects possibly include the fact that the retinal images in our dataset were taken from a rather homogeneous population base in terms of ethnicity; the technical quality and the rate of standardization in the imaging setup within our images, and the quality of their gradings may also have attributed to the goodness of results.

We also investigated the effect of the size of the images used in training, on the performance of the trained deep learning system in the fundus image classification, an assessment which has not been considered in the literature before. In our investigation, the classifier was trained on five different input image sizes, for each of the five classification systems. In all tasks, the best performing model according to AUC / macro-AUC metric was the model with the largest resolution, namely the images of $2095 \times 2095$ pixels, except that under the QRDR classification the best results were achieved with image sizes of $1024 \times 1024$ and 2095 x 2095 pixels.

It was observed that the AUC performance overall increased with the input image size, which could be attributed to the fact that the amount of information and features in the images increases with the image size. However, in our implementation, increasing the input image size without modifying the model architecture, dramatically increased the average duration of training and inference as well as memory requirements in the computations. This in fact meant that the modeling setup had to be different for the $2095 \times 2095$ input images than for the other ones. As an example, training a slightly modified model using the input size of $2095 \times 2095$ pixels with mini-batch size of 1 increased the average training time for each epoch approximately 20 times in comparison to the base model using an input size of $512 \times 512$ pixels with the mini-batch size of 6, but with minimal performance improvements in most classification tasks. This suggests that in the case of a fixed wall-clock time requirement for training and inference, it might be better to consider a smaller input image size, and do a wider (grid-)search in terms of the other hyper-parameters, possibly leading to better results, with the same wall-clock time.

Finally, we acknowledge the following limitations of our present deep learning AI-system. The first one concerns the image grading reference, which in our case was provided as an agreement of two experienced and qualified graders for each image, but could unavoidably include grader biases that can result in decreased generalization performance of the model. In addition, as was expressed in Gulshan et al.[4], deep learning neural networks have an inherent limitation of possibly learning features that are unknown or ignored by medical experts, when the network is only fed in an image and its grading without defining diagnostically important features such as *microaneurysms* and *exudates* as well as their numbers that are important biomarkers of diabetic retinopathy.

In our study we have demonstrated that a deep learning AI-system applied to a relatively small retinal image dataset could accurately identify the severity grades of diabetic retinopathy and macular edema and that its accuracy was improved by using high resolution and quality images.

## Code availability

Code used for preprocessing and deep learning includes proprietary parts and cannot be released publicly. However, preprocessing and deep learning algorithms can be replicated using the information in the Methods section and on the Supplementary information.

## Data availability

Datasets used in model training, tuning and primary validation were provided by Digifundus Ltd. This dataset is not publicly available and restriction apply to their use. The Messidor dataset may be requested from [http://www.adcis.net/en/Download-Third-Party/Messidor.html](http://www.adcis.net/en/Download-Third-Party/Messidor.html).

# Acknowledgements

**Author Affiliations:**

Jaakko Sahlsten (JS), Joel Jaskari (JJ), Jyri Kivinen (JK), Kimmo Kaski (KK);

Aalto University School of Science, Finland

Lauri Turunen (LT), Esa Jaanio (EJ);

Digifundus Ltd, Finland

Kustaa Hietala (KH);

Central Finland Central Hospital, Finland

**Author Contributions:**

JS & JJ & JK: research and experiment planning, retinal image preprocessing, deep learning neural network design, experiments, results analysis, and manuscript writing;

LT & EJ: research and experiment planning, retinal image grading, writing parts of manuscript;

KH: edited, revised manuscript;

KK: research and experiment planning, results analysis and interpretation, and manuscript writing, revisions, and final polishing

**Conflict of Interest Disclosures:**

JS, JJ, JK, KK, LT, EJ, KH: None

**Funding/Support:**

JS & JJ & JK: Digifundus Ltd

KH: Silmäsäätiö

**Role of the Funder/Sponsor:**

Digifundus Ltd: partial project funding; provider of the retinal image dataset (over 40000 images) and image grading by medical experts

Silmäsäätiö:

**Additional Contributions:**

None

# Figures

**Figure 1: ROC curves for non-referable vs. referable diabetic retinopathy in classifying non-referable vs. referable macular edema on primary validation set and Messidor set.**

**A: NRDR/RDR classification on the primary validation set (N=7118).**

**B: NRDR/RDR classification on Messidor set (N=1200).**

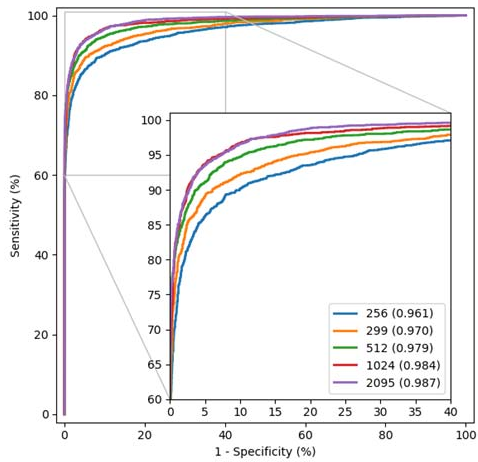
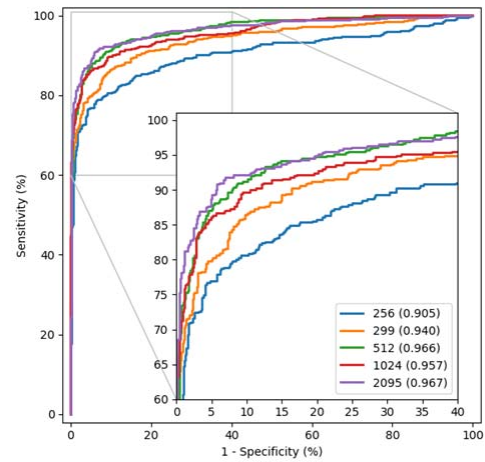

**C: NRDME/RDME classification on the primary validation set (N=7304).**

**D: NRDME/RDME classification on Messidor set (N=1200).**

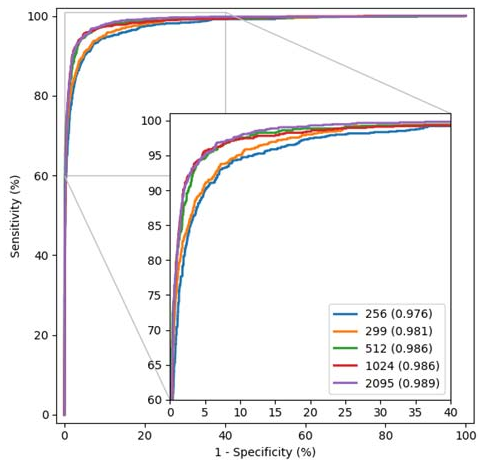
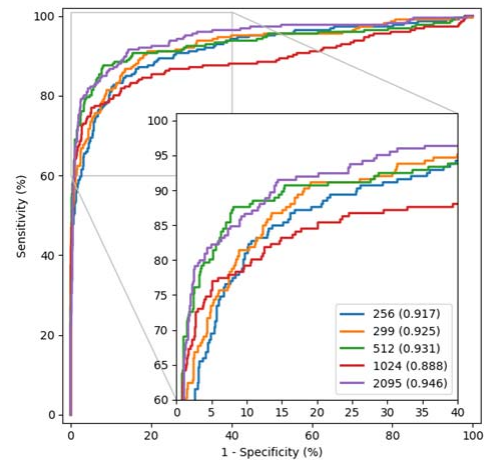

Referable vs. non-referable diabetic retinopathy shown in A and B and referable diabetic macular edema shown in C and D. ROC curve is shown for input image sizes of 256 × 256, 299 × 299, 512 × 512, 1024 × 1024 and 2095 × 2095 pixels. AUC shown in parentheses in the legend.

**Figure 2: ROC curves for best performing model for each of the multiclass classification tasks.**

**A: PIRC classification on the primary validation set (N=7129) with input size 2095 × 2095.**

**B: PIMEC classification on the primary validation set (N=7304) with input size 2095 × 2095.**

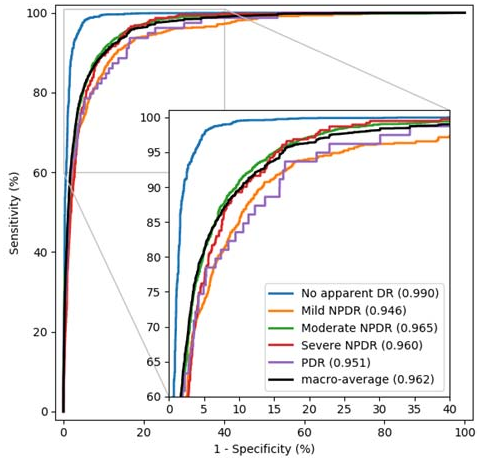
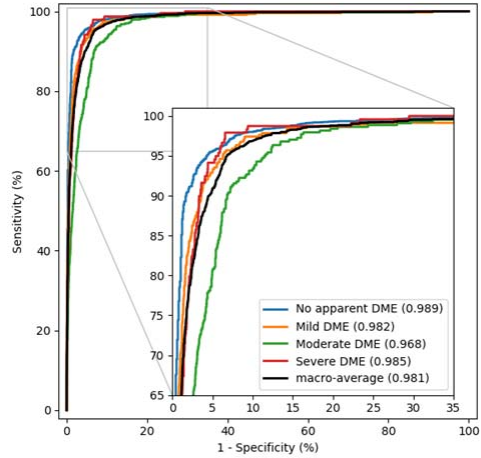

**C: QRDR classification on the primary validation set (N=8226) with input size 1024 × 1024.**

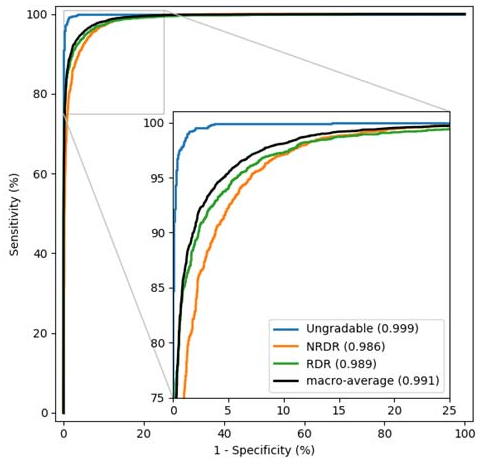

Multiclass tasks include PIRC, PIMEC and QRDR for the best performing models based on macro-AUC. ROC curves are shown for each class in one-vs-all strategy with addition of macro-average of ROC. Positive class marked in legend with AUC shown in parentheses.

# Tables

**Table 1: Summary (amounts, percentages) on the dataset divisions between training, tuning, and primary validation used in experiments.**

| Grading system | Patients / Label | Training | Tuning | Primary validation |
|---|---|---|---|---|
| RDR | Patients, No. | 8694 | 1313 | 2477 |
| | Images Total, No. (%) | 24806 (100) | 3706 (100) | 7118 (100) |
| | Images Grade 0, No. (%) | 13895 (56.0) | 2079 (56.1) | 4031 (56.6) |
| | Images Grade 1, No. (%) | 10911 (44.0) | 1627 (43.9) | 3087 (43.4) |
| PIRC | Patients, No. | 8770 | 1259 | 2455 |
| | Images Total, No. (%) | 24941 (100) | 3560 (100) | 7129 (100) |
| | Images Grade 0, No. (%) | 11160 (44.7) | 1573 (44.2) | 3229 (45.3) |
| | Images Grade 1, No. (%) | 2793 (11.2) | 408 (11.5) | 842 (11.8) |
| | Images Grade 2, No. (%) | 9221 (37.0) | 1312 (36.9) | 2597 (36.4) |
| | Images Grade 3, No. (%) | 1480 (5.9) | 225 (6.3) | 382 (5.4) |
| | Images Grade 4, No. (%) | 287 (1.2) | 42 (1.2) | 79 (1.1) |
| RDME | Patients, No. | 8669 | 1281 | 2534 |
| | Images Total, No. (%) | 24651 (100) | 3675 (100) | 7304 (100) |
| | Images Grade 0, No. (%) | 20819 (84.5) | 3113 (84.7) | 6162 (84.4) |
| | Images Grade 1, No. (%) | 3832 (15.5) | 562 (15.3) | 1142 (15.6) |
| PIMEC | Patients, No. | 8708 | 1242 | 2534 |
| | Images Total, No. (%) | 24791 (100) | 3535 (100) | 7304 (100) |
| | Images Grade 0, No. (%) | 20958 (84.5) | 2974 (84.1) | 6162 (84.4) |
| | Images Grade 1, No. (%) | 1531 (6.2) | 237 (6.7) | 465 (6.4) |
| | Images Grade 2, No. (%) | 1566 (6.3) | 222 (6.2) | 438 (6.0) |
| | Images Grade 3, No. (%) | 736 (3.0) | 102 (2.9) | 239 (3.3) |
| QRDR | Patients, No. | 10232 | 1466 | 2926 |
| | Images Total, No. (%) | 28787 (100) | 4109 (100) | 8226 (100) |
| | Images Grade 0, No. (%) | 3827 (13.3) | 533 (13.0) | 1132 (13.8) |
| | Images Grade 1, No. (%) | 14005 (48.7) | 1991 (48.5) | 4009 (48.7) |
| | Images Grade 2, No. (%) | 10955 (38.1) | 1585 (38.6) | 3085 (37.5) |

**Table 2: Comparison of classification results for referable diabetic retinopathy.**

| Author | Train samples | Validation samples | Input size | AUC | Sensitivity | Specificity |
|---|---|---|---|---|---|---|
| Gulshan et al.[4] | 118419 | 8788 | 299 | 0.991 (0.988-0.993)[a] | 0.903 (0.875-0.927)[a] | 0.981 (0.978-0.985)[a] |
| Ting et al.[5] | 76370 | 71896 | 512 | 0.936 (0.925-0.943)[b] | 0.905 (0.873-0.930)[c] | 0.916 (0.910-0.922)[c] |
| Ours | 28512 | 7118 | 2095 | 0.987 (0.984-0.989)[a] | 0.896 (0.885-0.907)[a] | 0.974 (0.969-0.979)[a] |

The train and validation samples refer to the image amounts in the respective sets, and the input size refers to the (input of the classifier network) image width and height in pixels. Train samples include the tuning set. Our operating point for sensitivity and specificity is calculated at 0.900 sensitivity for comparison of results at similar operating point to Gulshan et al.[4] and Ting et al.[5].

[a] 95% exact CI calculated with Clopper-Pearson method.
[b] 95% asymptotic, bias-corrected CI calculated with cluster-bootstrap on patient level.
[c] 95% asymptotic CI calculated for each logit with cluster sandwich using on patient level.

**Table 3: Classification results for PIRC, QRDR and PIMEC with varying input image sizes on the primary validation set.**

| Grading system | Input size | Macro-AUC | Accuracy | Quadratic-Weighted Kappa |
|---|---|---|---|---|
| PIRC | 256 | 0.901 | 0.751 | 0.772 |
| PIRC | 299 | 0.919 | 0.785 | 0.834 |
| PIRC | 512 | 0.951 | 0.838 | 0.894 |
| PIRC | 1024 | 0.961 | **0.870** | **0.915** |
| PIRC | 2095* | **0.962** | 0.869 | 0.910 |
| QRDR | 256 | 0.977 | 0.912 | 0.901 |
| QRDR | 299 | 0.981 | 0.922 | 0.914 |
| QRDR | 512 | 0.989 | 0.937 | 0.930 |
| QRDR | 1024 | **0.991** | **0.938** | **0.932** |
| QRDR | 2095* | **0.991** | 0.925 | 0.914 |
| PIMEC | 256 | 0.959 | 0.928 | 0.813 |
| PIMEC | 299 | 0.970 | 0.923 | 0.803 |
| PIMEC | 512 | 0.979 | 0.935 | 0.832 |
| PIMEC | 1024 | 0.978 | **0.937** | 0.846 |
| PIMEC | 2095* | **0.981** | 0.934 | **0.856** |

Macro-AUC refers to area under macro average of ROC for each class one-vs-all manner. 'Input size' refers to the heights and widths of the input images in pixels.
* Model trained using instance normalization layers, instead of batch normalization, and optimizer updates accumulated over 15 mini-batches.

**Table 4: Classification results for predicted of NRDR/RDR and NRDME/RDME with varying input image sizes on the primary validation set.**

| Grading system | Input size | AUC | Sensitivity | Specificity | Accuracy |
|---|---|---|---|---|---|
| RDR | 256 | 0.961 (0.956-0.965) | 0.895 (0.884-0.906) | 0.913 (0.904-0.921) | 0.905 (0.898-0.912) |
| RDR | 299 | 0.970 (0.966-0.974) | 0.896 (0.884-0.906) | 0.946 (0.938-0.953) | 0.924 (0.918-0.930) |
| RDR | 512 | 0.979 (0.975-0.982) | 0.900 (0.888-0.910) | 0.963 (0.956-0.968) | 0.935 (0.929-0.941) |
| RDR | 1024 | 0.984 (0.981-0.987) | **0.910 (0.899-0.920)** | 0.970 (0.964-0.975) | **0.944 (0.938-0.949)** |
| RDR | 2095* | **0.987 (0.984-0.989)** | 0.896 (0.885-0.907) | **0.974 (0.969-0.979)** | 0.940 (0.935-0.946) |
| RDME | 256 | 0.976 (0.973-0.980) | 0.891 (0.871-0.908) | 0.953 (0.948-0.958) | 0.943 (0.938-0.949) |
| RDME | 299 | 0.981 (0.977-0.984) | 0.891 (0.871-0.908) | 0.960 (0.955-0.965) | 0.949 (0.944-0.954) |
| RDME | 512 | 0.986 (0.983-0.989) | 0.890 (0.870-0.907) | 0.976 (0.972-0.980) | 0.963 (0.958-0.967) |
| RDME | 1024 | 0.986 (0.983-0.989) | **0.921 (0.904-0.936)** | 0.974 (0.970-0.978) | 0.966 (0.961-0.970) |
| RDME | 2095* | **0.989 (0.986-0.991)** | 0.905 (0.887-0.922) | **0.978 (0.974-0.982)** | **0.967 (0.963-0.971)** |

Sensitivity, specificity and accuracy measured at 0.900 sensitivity operating point of tuning set. 95% exact Clopper-Pearson confidence interval in brackets. 'input size' refers to the heights and widths of the input images in pixels.
* Trained with model using instance normalization layers and an optimizer with accumulation of 15 mini-batches.

**Table 5: Classification results for predicted of RDR and RDME with varying input image sizes on the Messidor dataset.**

| Grading system | Input size | AUC | Sensitivity | Specificity | Accuracy |
|---|---|---|---|---|---|
| RDR | 256 | 0.905 (0.887-0.921) | 0.826 (0.790-0.858) | 0.872 (0.845-0.896) | 0.853 (0.831-0.872) |
| RDR | 299 | 0.940 (0.926-0.953) | 0.906 (0.877-0.930) | 0.824 (0.794-0.852) | 0.859 (0.838-0.878) |
| RDR | 512 | 0.966 (0.954-0.976) | **0.945 (0.922-0.963)** | 0.811 (0.780-0.840) | 0.868 (0.848-0.887) |
| RDR | 1024* | 0.957 (0.944-0.968) | 0.853 (0.820-0.883) | 0.955 (0.937-0.969) | 0.912 (0.894-0.927) |
| RDR | 2095*,** | **0.967 (0.955-0.976)** | 0.859 (0.826-0.888) | **0.971 (0.956-0.982)** | **0.923 (0.907-0.938)** |
| RDME | 256 | 0.917 (0.900-0.932) | 0.619 (0.553-0.683) | 0.969 (0.956-0.979) | 0.903 (0.885-0.919) |
| RDME | 299 | 0.925 (0.908-0.939) | 0.633 (0.566-0.696) | 0.975 (0.964-0.984) | 0.911 (0.893-0.926) |
| RDME | 512 | 0.931 (0.915-0.944) | **0.690 (0.626-0.750)** | 0.989 (0.980-0.994) | **0.932 (0.917-0.946)** |
| RDME | 1024* | 0.888 (0.869-0.905) | 0.606 (0.539-0.670) | 0.991 (0.983-0.996) | 0.918 (0.901-0.933) |
| RDME | 2095*,** | **0.946 (0.932-0.958)** | 0.597 (0.530-0.662) | **0.992 (0.984-0.996)** | 0.917 (0.900-0.932) |

Classification on the Messidor set[14]. Sensitivity, specificity and accuracy measured at 0.900 sensitivity operating point of tuning set. 95% exact Clopper-Pearson confidence interval in brackets. 'input size' refers to the heights and widths of the input images in pixels.
* Trained with model using instance normalization layers and an optimizer with accumulation of 15 mini-batches.
** Messidor images upscaled from input image size of 900 × 900 pixels using bicubic interpolation

**Table 6: Confusion matrices for PIRC, PIMEC and QRDR classification tasks with varying input size on the primary validation set.**

| Input size | | PIRC | | | | | | PIMEC | | | | | QRDR | | |
|---|---|---|---|---|---|---|---|---|---|---|---|---|---|---|---|
| **256** | | 0 | 1 | 2 | 3 | 4 | | 0 | 1 | 2 | 3 | | 0 | 1 | 2 |
| | 0 | 2982 | 147 | 99 | 0 | 1 | 0 | 6053 | 49 | 38 | 22 | 0 | 1084 | 45 | 3 |
| | 1 | 500 | 250 | 92 | 0 | 0 | 1 | 98 | 332 | 32 | 3 | 1 | 18 | 3736 | 255 |
| | 2 | 343 | 187 | 2040 | 27 | 0 | 2 | 80 | 61 | 259 | 38 | 2 | 0 | 401 | 2684 |
| | 3 | 13 | 8 | 279 | 82 | 0 | 3 | 30 | 10 | 66 | 133 | | | | |
| | 4 | 3 | 0 | 51 | 24 | 1 | | | | | | | | | |
| **299** | | 0 | 1 | 2 | 3 | 4 | | 0 | 1 | 2 | 3 | | 0 | 1 | 2 |
| | 0 | 3027 | 142 | 60 | 0 | 0 | 0 | 5977 | 79 | 70 | 36 | 0 | 1093 | 38 | 1 |
| | 1 | 438 | 283 | 121 | 0 | 0 | 1 | 64 | 353 | 42 | 6 | 1 | 35 | 3717 | 257 |
| | 2 | 197 | 198 | 2128 | 74 | 0 | 2 | 63 | 45 | 295 | 35 | 2 | 0 | 311 | 2774 |
| | 3 | 6 | 5 | 215 | 154 | 2 | 3 | 21 | 9 | 95 | 114 | | | | |
| | 4 | 4 | 0 | 45 | 25 | 5 | | | | | | | | | |
| **512** | | 0 | 1 | 2 | 3 | 4 | | 0 | 1 | 2 | 3 | | 0 | 1 | 2 |
| | 0 | 3178 | 36 | 15 | 0 | 0 | 0 | 6015 | 51 | 78 | 18 | 0 | 1108 | 22 | 2 |
| | 1 | 296 | 440 | 104 | 2 | 0 | 1 | 60 | 374 | 29 | 2 | 1 | 50 | 3786 | 173 |
| | 2 | 114 | 201 | 2088 | 193 | 1 | 2 | 57 | 49 | 317 | 15 | 2 | 1 | 271 | 2813 |
| | 3 | 3 | 3 | 119 | 256 | 1 | 3 | 19 | 8 | 91 | 121 | | | | |
| | 4 | 4 | 0 | 21 | 40 | 14 | | | | | | | | | |
| **1024** | | 0 | 1 | 2 | 3 | 4 | | 0 | 1 | 2 | 3 | | 0 | 1 | 2 |
| | 0 | 3106 | 95 | 28 | 0 | 0 | 0 | 6078 | 45 | 29 | 10 | 0 | 1115 | 16 | 1 |
| | 1 | 149 | 579 | 111 | 0 | 3 | 1 | 63 | 364 | 36 | 2 | 1 | 78 | 3732 | 199 |
| | 2 | 54 | 195 | 2257 | 89 | 2 | 2 | 70 | 55 | 268 | 45 | 2 | 0 | 218 | 2867 |
| | 3 | 4 | 2 | 138 | 222 | 16 | 3 | 30 | 6 | 67 | 136 | | | | |
| | 4 | 4 | 0 | 16 | 18 | 41 | | | | | | | | | |
| **2095*** | | 0 | 1 | 2 | 3 | 4 | | 0 | 1 | 2 | 3 | | 0 | 1 | 2 |
| | 0 | 3185 | 27 | 16 | 1 | 0 | 0 | 6024 | 68 | 51 | 19 | 0 | 1069 | 63 | 0 |
| | 1 | 169 | 517 | 156 | 0 | 0 | 1 | 42 | 388 | 29 | 6 | 1 | 22 | 3926 | 61 |
| | 2 | 81 | 143 | 2304 | 69 | 0 | 2 | 47 | 69 | 262 | 60 | 2 | 1 | 472 | 2612 |
| | 3 | 3 | 1 | 188 | 189 | 1 | 3 | 14 | 6 | 71 | 148 | | | | |
| | 4 | 4 | 0 | 25 | 47 | 3 | | | | | | | | | |

Ground truth shown in rows and predicted classes in columns. PIRC classes (0 = no apparent DR, 1 = mild NPDR, 2 = moderate NPDR, 3 = severe NPDR, 4 = PDR), PIMEC classes (0 = no apparent DME, 1 = mild DME, 2 = moderate DME, 3 = severe DME) and QRDR classes (0 = ungradable, 1 = NRDR, 2 = RDR)

* Model trained using instance normalization layers, instead of batch normalization, and optimizer updates accumulated over 15 mini-batches.

# References

1. Marshall SM, Flyvbjerg A. Prevention and early detection of vascular complications of diabetes. *BMJ.* 2006;333(7566):475-480.
2. Hutchinson A, McIntosh A, Peters J, et al. Effectiveness of screening and monitoring tests for diabetic retinopathy--a systematic review. *Diabet Med.* 2000;17(7):495-506.
3. Taylor R, Batey D. *Handbook of Retinal Screening in Diabetes: Diagnosis and Management.* Wiley; 2012.
4. Gulshan V, Peng L, Coram M, et al. Development and Validation of a Deep Learning Algorithm for Detection of Diabetic Retinopathy in Retinal Fundus Photographs. *JAMA.* 2016;316(22):2402-2410.
5. Ting DSW, Cheung CY, Lim G, et al. Development and Validation of a Deep Learning System for Diabetic Retinopathy and Related Eye Diseases Using Retinal Images From Multiethnic Populations With Diabetes. *JAMA.* 2017;318(22):2211-2223.
6. Wilkinson CP, Ferris FL, 3rd, Klein RE, et al. Proposed international clinical diabetic retinopathy and diabetic macular edema disease severity scales. *Ophthalmology.* 2003;110(9):1677-1682.
7. Summanen P, Kallioniemi V, Komulainen J, et al. [Update on Current Care Guideline: Diabetic retinopathy]. *Duodecim.* 2015;131(9):893-894.
8. Krause J, Gulshan V, Rahimy E, et al. Grader Variability and the Importance of Reference Standards for Evaluating Machine Learning Models for Diabetic Retinopathy. Ophthalmology. 2018;125(8):1264-1272.
9. Guan MY, Gulshan V, Dai AM, Hinton GE. Who Said What: Modeling Individual Labelers Improves Classification. *arXiv e-prints.* 2017. https://ui.adsabs.harvard.edu/\#abs/2017arXiv170308774G. Accessed March 01, 2017.
10. Wang F, Casalino LP, Khullar D. Deep Learning in Medicine-Promise, Progress, and Challenges. *JAMA Intern Med.* 2018.
11. Medical Research Act. 2010; https://www.finlex.fi/en/laki/kaannokset/1999/en19990488_20100794.pdf.
12. Goodfellow I, Bengio Y, Courville A. Deep Learning. MIT Press; 2016.
13. Szegedy C, Vanhoucke V, Ioffe S, Shlens J, Wojna Z. Rethinking the Inception Architecture for Computer Vision. *arXiv e-prints.* 2015. https://ui.adsabs.harvard.edu/\#abs/2015arXiv151200567S. Accessed December 01, 2015.
14. Russakovsky O, Deng J, Su H, et al. ImageNet Large Scale Visual Recognition Challenge. 2015;115(3):211-252.
15. Decencière E, Zhang X, Cazuguel G, et al. *Feedback on a publicly distributed image database: The Messidor database.* Vol 02014.

# Supplementary Information

## Experimental setup

Initially the network architecture was that of the Inception-v3 optimized for ImageNet dataset[1] classification, but with the exception that the fully-connected layer of the neural network model was replaced with two consecutive fully-connected layers, the former utilizing a regularization technique called dropout and the latter a vanilla fully-connected layer with softmax nonlinearity to define the diabetic retinopathy grade probabilities. Values of all other model parameters than those of the added fully-connected layers were initialized to the parameter values of the model (pre)trained on the ImageNet dataset, and all of the parameters were updated during the training.

Network parameters were fine-tuned with the Adam algorithm[2], also called Adam optimizer, on input image sizes of 2095 × 2095, 1024 × 1024, 512 × 512, 299 × 299, and 256 × 256 pixels. The hyperparameters including the learning rate, dropout rate and mini-batch size were tuned on the tuning set.

The networks using the input image sizes of 2095 × 2095 pixels were defined and trained using minor modifications to the training procedure and to the network architecture. They were trained with the mini-batch size 1, due to memory restrictions of Graphical Processing Units (GPUs) used deep learning neural network computations. The canonical model and optimizer were modified to take deviation into consideration as follows: the batch normalization layers were replaced with instance normalization and the optimizer updates were accumulated and averaged to attain similar updates as with larger mini-batch size. The networks for the input image size of 2095 × 2095 pixels were also trained only with the best guess estimates of the appropriate hyperparameter values due to time restrictions.

The early stopping approach was used in the hyperparameter search with the stopping criterion chosen as the area under the receiver operating characteristic curve (AUC) on a binary classification task and the area under the average of the receiver operating characteristic curves of each class in one-to-all manner (macro-AUC) on a multi-class classification task. In addition, the learning rate was set to decay exponentially so that the learning rate at an epoch $\tau \in [2, \ldots]$ was $0.99^{\tau-1}$ times the learning rate at the first epoch ($\tau = 1$), the initial learning rate. Our experiments were constructed with Keras and TensorFlow deep learning frameworks.

## Supplementary References